\documentclass[manuscript,screen]{acmart}

\usepackage{caption}
\usepackage{subcaption}
\usepackage{bm}
\usepackage{mathtools, nccmath}

\AtBeginDocument{%
  \providecommand\BibTeX{{%
    \normalfont B\kern-0.5em{\scshape i\kern-0.25em b}\kern-0.8em\TeX}}}

\setcopyright{rightsretained}
\copyrightyear{2020}
\acmYear{2020}
\acmDOI{xx.xxxx/xxxxxxx.xxxxxxx}

\acmConference[RecSys '20]{RecSys '20: CARS 2.0: Workshop on Context-Aware Recommender Systems}{September 22--26, 2020}{CARS Workshop, Online}
\acmBooktitle{acmBooktitle}
\acmPrice{15.00}
\acmISBN{978-1-4503-XXXX-X/18/06}

\begin{document}

\title[Contextual Re-ranking of Music Recommendations]{Contextual Personalized Re-Ranking of Music Recommendations through Audio Features}

\author{Boning Gong}
\email{b.gong@student.tudelft.nl}
\affiliation{%
  \institution{Delft University of Technology}
  \streetaddress{Van Mourik Broekmanweg 6}
  \city{Delft}
  \postcode{2628 XE}
  \country{The Netherlands}
}

\author{Mesut Kaya}
\email{m.kaya@tudelft.nl}
\affiliation{%
  \institution{Delft University of Technology}
  \streetaddress{Van Mourik Broekmanweg 6}
  \city{Delft}
  \postcode{2628 XE}
  \country{The Netherlands}
}

\author{Nava Tintarev}
\email{n.tintarev@tudelft.nl}
\affiliation{%
  \institution{Delft University of Technology}
  \streetaddress{Van Mourik Broekmanweg 6}
  \city{Delft}
  \postcode{2628 XE}
  \country{The Netherlands}
}

\renewcommand{\shortauthors}{Gong, et al.}

\begin{abstract}
Users are able to access millions of songs through music streaming services like Spotify, Pandora, and Deezer. Access to such large catalogs, created a need for relevant song recommendations. However, user preferences are highly subjective in nature and change according to context (e.g., music that is suitable in the morning is not as suitable in the evening). Moreover, the music one user may prefer in a given context may be different from what another user prefers in the same context (i.e., what is considered good morning music differs across users). Accurately representing these preferences is essential to creating accurate and effective song recommendations. User preferences for songs can be based on high level audio features, such as tempo and valence. In this paper, we therefore propose a contextual re-ranking algorithm, based on \textit{audio feature} representations of user preferences in specific contextual conditions. We evaluate the performance of our re-ranking algorithm using the \#NowPlaying-RS dataset, which exists of user listening events crawled from Twitter and is enriched with song audio features. We compare a global (context for all users) and personalized (context for each user) model based on these audio features. The global model creates an audio feature representation of each contextual condition based on the preferences of all users. Unlike the global model, the personalized model creates \textit{user-specific} audio feature representations of contextual conditions, and is measured across 333 distinct users. We show that the personalized model outperforms the global model when evaluated using the precision and mean average precision metrics. 
\end{abstract}




\maketitle

\section{Introduction}\label{section:introduction}
Users can utilize streaming services like Spotify, Pandora and Deezer to access tens of millions of songs from all around the world.\footnote{https://www.businessofapps.com/data/spotify-statistics/\#4, retrieved July 2020} Music recommender systems assist users in finding and discovering songs by filtering the most relevant ones. To create accurate recommendations, contextual information of the users is important, because of its influence on people's short term music preferences~\cite{north2004uses}. However, music preferences of users change through influences from the physical environment, such as activities or geo-location~\cite{kaminskas2012contextual}. People prefer different music when working out in a gym compared to reading a book on the couch, for example. One way to improve personalization is to incorporate this contextual data of the user~\cite{adomavicius2005incorporating}. This led to an emerging interest in context-aware music recommender systems~\cite{braunhofer2013location}.

We use two definitions to describe context. The first are contextual dimensions, which are categories of contexts, e.g. \textit{time of day}, \textit{activity} etc. The second are contextual conditions. A contextual dimension is made up of multiple contextual conditions, e.g. \textit{morning} and \textit{afternoon} within \textit{time of day}.

Recommendation systems that use contextual information can be divided into 3 categories, namely contextual pre-filtering, contextual post-filtering, and contextual modeling~\cite{adomavicius2011context}. Contextual modeling is the most powerful method of the three~\cite{zheng2018context}. These techniques have been applied to capture the relation between contexts and user preferences. However, these models are often hard to understand, which makes it hard to explain the recommendations to users~\cite{zhang2019deep}. Both filtering methods have the benefit over contextual modeling that no additional changes are required to the existing recommender system. For pre-filtering, only the input is adjusted, while for post-filtering the resulting recommendation is altered~\cite{villegas2018characterizing}. Pre-filtering methods have been well developed, but post-filtering research efforts have been limited~\cite{zheng2018context}.

In this paper, we propose a post-filtering approach \footnote{It might be considered a hybrid model in case the output of another context-aware recommender is used.}: a contextual re-ranking algorithm, which ranks higher songs that are more suitable to a user's current contextual condition. It can be applied to any existing music recommender's output, is easily understandable and explainable, and works with any contextual dimension. Users like or dislike songs based on the characteristics of their audio  features~\cite{wang2014improving}. Significant correlations exist between music preferences expressed in audio features and personality traits~\cite{10.1145/3340631.3394874}. Therefore, we use audio features to model user preferences for specific contextual conditions. To the best of our knowledge, our approach is novel, since it uses audio features to models users' context specific preferences when re-ranking songs. Music preference is also highly subjective; while one person may experience two songs as dissimilar, a second one may feel a high resemblance~\cite{schedl2013neglected}. What is suitable in a given context may therefore also differ from person to person. That is why we compare a \textit{global} preference model with \textit{personalized} user preference models for the re-ranking algorithm. Whereas the \textit{personalized} model is based on \underline{individual} preferences in different contextual conditions, the \textit{global} model is based on context preferences of \underline{all} users. Our work addresses the following research questions:

\begin{itemize}
    \item \textbf{RQ1:} How are \textit{contextual conditions} of different contextual dimensions related to \textit{audio features}? 
    \item \textbf{RQ2:} How does \textit{re-ranking}, based on audio feature representations of user preferences in different contextual conditions, affect music \textit{recommendation accuracy}?
    \item \textbf{RQ3:} How do \textit{global} audio feature representations of user preferences in different contextual conditions affect the re-ranking results compared to \textit{personalized} audio feature representations of user preferences in the same contextual condition (time of day)?
\end{itemize} 

The rest of this paper is organized as follows. First, a description of the related work is given. The analysis of the relation between audio features and context is described in Section~\ref{section:context_audio_feature_analysis}, and it aims to answer RQ1. Section~\ref{section:reranking_algorithm} elaborates on the proposed re-ranking algorithm and the two user preference models that are created to address RQ2 and RQ3. The implementation, evaluation, and results of the re-ranking algorithm are given in Section~\ref{section:experiment} before concluding in Section~\ref{section:conclusion}. 

\section{Related Work}\label{section:related_work}
This section begins with a summary of work on context-aware recommender systems with a focus on \textit{post-filtering} techniques. The second section discusses music recommender systems where \textit{audio features} have been used. We conclude with outlining the novelty of our approach relative to the state-of-the-art.
%~\ref{subsection:audio_features_recommender_systems}

\subsection{Context-Aware Recommender Systems}\label{subsection:context_aware_recommender_systems}
Context-aware recommender systems extend traditional recommender systems by taking information of users' contextual situation into account. Here, context is defined as any information that can be used to characterize the situation of users (e.g. location, activity, weather, mood etc.) that are considered relevant to the interaction between a user and an application. Adomavicius et al. were one of the first to use such information in recommender systems~\cite{adomavicius2005incorporating}. Quality of recommendations has shown to be improved through using contextual information by multiple researchers~\cite{villegas2018characterizing}. 

Contextual post-filtering approaches apply context-dependent factors to the list of recommendations, which are given by a traditional recommender algorithm (e.g., matrix factorization). The order of the songs in the given recommendation list are adjusted to the given context~\cite{haruna2017comprehensive}. This allows usage of traditional recommender algorithms, without the need to change them. Panniello et al. proposed two contextual post-filtering approaches, which they call \emph{Weight PoF} and \emph{Filter PoF}~\cite{panniello2009experimental}. \emph{Filter PoF} removes items that are least relevant to the given context, while \emph{Weight PoF} reorders items based on the rating probability of relevance in the given context. The probabilities are created based on the behavior of most similar users in the same context. Cremonesi et al. use association rules mining between item characteristics and contextual knowledge to find correlations~\cite{cremonesi2011top}. A subset of items is selected based on their correlation from the initial recommendation list and the top-N within this subset is recommended. Lamche et al. use their own distance metric to calculate the similarity between the user's current context and an item's representative context~\cite{lamche2015context}. 

\subsection{Audio Features in Recommender Systems}\label{subsection:audio_features_recommender_systems}
Cheng and Shen created Just-for-me which uses a unified probabilistic generative model to model audio features and context in a latent space~\cite{cheng2014just}. Songs are represented by 3 different acoustic features, which are measured using 0.5s frame sequences. Chen et al. analyzed the relation between emotions, through user-generated text, and
music through factorization machines~\cite{chen2013using}. They embedded audio features, including loudness, mode, tempo, and danceability that were extracted using the EchoNest API. Schedl et al. combined music context and music content in a hybrid model~\cite{schedl2014location}. The audio features they used include onset patterns and coefficients, timbral features, and two custom descriptors for attackness and harmonicness. Song similarities were estimated using these features and used to generate recommendations. \\

\noindent \textit{Novelty.} Our work belongs to the group of contextual post-filtering approaches. Similarly to previous approaches, our approach uses a similar weighing function as \emph{Weight PoF}~\cite{panniello2009experimental}. However, instead of using a rating probability of relevance based on similar users, we use context specific audio feature representations to measure similarity. Where Lamche et al.~\cite{lamche2015context} create context models around items, we do the opposite by creating audio feature models around contexts. Furthermore, unlike~\cite{cheng2014just,chen2013using}, our representation does not use any matrix factorization techniques or create any other latent spaces. Instead, we use a simple vector representation which allows for straightforward distance measurements when comparing songs to contexts. Our novel approach gives an interesting performance comparison to the more complex latent models.

\section{Context-Audio Feature Analysis}\label{section:context_audio_feature_analysis}
In this section, we discuss the analysis we carried out to answer RQ1.\footnote{The code and all results can be found at https://github.com/boninggong/ContextAudioFeaturesAnalysis} The results give us an idea of how valuable audio features are in representing user preferences in different contextual conditions. 

\subsection{Audio Features}\label{subsection:audio_features}
Through the Spotify API,\footnote{\url{https://developer.spotify.com/documentation/web-api/}, retrieved June 2020} developers can retrieve a variety of audio features for any given song that is available on Spotify, the music streaming platform. This audio features endpoint provides high-level acoustic attributes based on the audio of a given song. Some features, like \textit{tempo}, are well-known, while others, like \textit{danceability}, are more specialized.\footnote{\url{https://developer.spotify.com/community/news/2016/03/29/audio-features-recommendations-user-taste/}, retrieved July 2020}

All audio features are precise for a given song and have values between 0 and 1, except for \emph{loudness} and \emph{tempo}. We normalize \emph{tempo}, whose values range from 0 to 220, by dividing the values by 220 and \emph{loudness}, whose values range from -40 to 0, by adding 40 before dividing by 40. Each audio feature has its own distribution\footnote{The distributions be found at \url{https://developer.spotify.com/documentation/web-api/reference/tracks/get-audio-features/}, retrieved June 2020.}, which might affect its distinctiveness and correlation with contextual conditions. We use them as is, to keep all audio features as close as possible to their original values. A total of 11 audio features are obtainable of which we include the following in our analysis: \textit{acousticness} (how acoustic a song is), \textit{danceability} (how suitable a song is for dancing), \textit{energy} (representing the activity and intensity of a song), \textit{instrumentalness} (including vocals or not), \textit{liveness} (whether the song has been recorded in a live setting), \textit{loudness} (physical strength/amplitude of a song), \textit{speechiness} (presence of spoken words), \textit{valence} (how much positivity a song contains), and \textit{tempo} (beats per minute, indicating the speed or pace of a song). \textit{Key} and \textit{mode} are represented by a limited number of non continuous values, which makes them unsuitable for our analysis.

\subsection{Contextual Dimensions and Conditions}\label{subsection:contextual_dimensions_conditions}
We use the contextual dimensions of \textit{activity} (\textit{running}, \textit{walking}, \textit{sleeping} and \textit{focusing}), \textit{time of day} (\textit{morning}, \textit{afternoon}, \textit{evening} and \textit{night}), and \textit{mood} (\textit{happy} and \textit{sad}). The reasons for using these dimensions are threefold. First, previous research has shown that they affect user preferences~\cite{andjelkovic2016moodplay,baltrunas2009towards,dias2014user}. Second, the conditions within
these dimensions are straightforward and there are many playlists on Spotify representing these conditions with
thousands of followers. Third, these dimensions are also present in available contextual music datasets, which we will use later-on, so the results are directly relevant to us.

\subsection{Analysis}\label{subsection:analysis}
We gather representative songs for each contextual condition through public playlists on Spotify. Pichl et al. showed that public playlists on Spotify represent different user preferences that are dependent on the \emph{intended use} or \emph{mood}~\cite{pichl2016understanding}. Furthermore, Cunningham et al. showed that people create and use playlists for specific contextual conditions~\cite{cunningham2006more}.  For each condition, at least 500 songs, from at least 4 different playlists created specific for that condition (e.g. ``Songs for sleeping''), were gathered. Each playlist that has been used has at least 1000 followers to make sure that multiple users agree on the quality of the playlist. For each song, we extract the audio features through the Spotify API. We average the values of these audio features for each condition and visualize them.

For each possible pair of conditions within a dimension and each audio feature, we carry out independent t-tests. The results of these tests tell us whether the differences might have happened by chance or whether they are significantly different. Two resulting values are especially important, the \textbf{t-score} and \textbf{p-value}. A t-score of 0 indicates two identical groups. The higher this score, the more different the two groups are. The p-value represents the probability of the results happening by chance and is always between 0 and 1. We apply Bonferroni correction, since we apply many independent t-tests. \footnote{There are a total of 117 tests, so we divide the usual p-value significance threshold of 0.05 by 117, resulting in a p-value threshold of 0.000427.}

\subsection{Results}\label{subsection:correlation_results}
A selection of the t-test results, is shown in Tables~\ref{table:afternoon-night} (afternoon--night),~\ref{table:running-relaxing} (running--relaxing),  and~\ref{table:happy-sad} (happy--sad). The complete results can be found in an external appendix. \footnote{\url{https://github.com/boninggong/ContextAudioFeaturesAnalysis}} Looking across all results, \emph{Liveness}, generally, is the weakest audio feature, but still reasonable in some comparisons. The other audio features are potentially good descriptors. 

The degree to which they correlate is both dimension and condition dependent. The resulting t-values tell us that some correlations are strongly positive, while others are negative. A majority of p-values, highlighted in bold, are below our Bonferroni corrected threshold, which means those are significant and have a very small chance of happening by chance. We thus conclude that certain audio features make strong descriptors to distinguish contextual conditions. In the next section we describe how we use audio features to represent conditions and use this in the re-ranking algorithm. 

Visualization of each condition in all 3 dimensions using radar and line plots can be found in the external appendix. \footnote{\url{https://github.com/boninggong/ContextAudioFeaturesAnalysis}} 

\begin{table}[ht]
    \begin{minipage}[t]{.26\textwidth}
    \captionof{table}{T-test results comparing the afternoon-night conditions.} \label{table:afternoon-night}
    \resizebox{\linewidth}{!}{
        \begin{tabular}{lrr}
        \toprule
        \textbf{afternoon-night} &      t &       p \\
        \midrule
                    acousticness &  -3.47 &  0.0005 \\
                    \textbf{danceability} &   \textbf{6.05} &  \textbf{0.0000} \\
                    \textbf{energy} &   \textbf{5.24} &  \textbf{0.0000} \\
                    \textbf{instrumentalness} & \textbf{-24.94} &  \textbf{0.0000} \\
                    liveness &   0.53 &  0.5986 \\
                    \textbf{loudness} &  \textbf{13.04} &  \textbf{0.0000} \\
                    \textbf{speechiness} &   \textbf{4.00} &  \textbf{0.0001} \\
                    tempo &   1.96 &  0.0507 \\
                    \textbf{valence} &  \textbf{14.12} &  \textbf{0.0000} \\
        \bottomrule
        \end{tabular}
    }
    \end{minipage}
    \hfill%
    \begin{minipage}[t]{.26\textwidth}
    \captionof{table}{T-test results comparing the running-relaxing conditions.} \label{table:running-relaxing}
    \resizebox{\linewidth}{!}{
        \begin{tabular}{lrr}
        \toprule
        \textbf{running-relaxing} &      t &       p \\
        \midrule
                    \textbf{acousticness} & \textbf{-45.53} &  \textbf{0.0000} \\
                    \textbf{danceability} &  \textbf{24.24} &  \textbf{0.0000} \\
                    \textbf{energy} &  \textbf{49.48} &  \textbf{0.0000} \\
                    \textbf{instrumentalness} & \textbf{-17.03} &  \textbf{0.0000} \\
                    \textbf{liveness} &   \textbf{9.13} &  \textbf{0.0000} \\
                    \textbf{loudness} &  \textbf{28.69} &  \textbf{0.0000} \\
                    \textbf{speechiness} &  \textbf{12.03} &  \textbf{0.0000} \\
                    \textbf{tempo} &  \textbf{10.40} &  \textbf{0.0000} \\
                    \textbf{valence} &  \textbf{23.95} &  \textbf{0.0000} \\
        \bottomrule
        \end{tabular}
    }
    \end{minipage}
    \hfill
    \begin{minipage}[t]{.26\textwidth}
    \captionof{table}{T-test results comparing the happy-sad conditions.} \label{table:happy-sad}
    \resizebox{\linewidth}{!}{
    \begin{tabular}{lrr}
    \toprule
    \textbf{happy-sad} &      t &       p \\
    \midrule
                \textbf{acousticness} & \textbf{-25.56} &  \textbf{0.0000} \\
                \textbf{danceability} &  \textbf{14.76} &  \textbf{0.0000} \\
                \textbf{energy} &  \textbf{26.92} &  \textbf{0.0000} \\
                instrumentalness &  -1.28 &  0.2026 \\
                \textbf{liveness} &   \textbf{6.04} &  \textbf{0.0000} \\
                \textbf{loudness} &  \textbf{18.52} &  \textbf{0.0000} \\
                \textbf{speechiness} &   \textbf{7.14} &  \textbf{0.0000} \\
                tempo &   0.71 &  0.4784 \\
                valence &  22.37 &  0.2531 \\
    \bottomrule
    \end{tabular}
    }
    \end{minipage}
\vspace{-12pt}
\end{table}
\section{Proposed Re-Ranking Algorithm}\label{section:reranking_algorithm}
In this section, first, we present a global and personalized model to model user preferences in contextual conditions using audio features. Thereafter, we elaborate on the re-ranking score calculation and briefly on an opposite variation based on this calculation.  

By way of notation, let $U$ be the set of all users and $S$ be the set of all songs. Let $c =\{c_1, c_2, \ldots, c_m\}$ be the set of contextual conditions. Let also $\Vec{s} = [a_1, a_2, \ldots, a_n]$ be the audio feature vector of a song $s \in S$, where $a_n$ is the value of the audio feature $n$ for song $s$. 

\subsection{Global model}\label{subsection:global_model}
The global model represents context specific user preferences through a vector of audio feature values, which are collected from all users, $U$. It uses all available positive user interactions in a given dataset to represent user preferences for different contextual conditions. The average of all audio feature values of the positively interacted songs will then form the representation. Thus, the global model can be represented as follows:

\begin{equation}
    \Vec{GM_{c_{k}}} = \left[a_1, a_2, \ldots, a_n\right] = \frac{1}{|S_{c_k}|}\cdot \sum_{s_j \in S_{c_k}} \Vec{s_j},
    \label{eq:global_model_formula}
\end{equation}

\noindent where $\Vec{GM_{c_{k}}}$ is a vector representing the global model for contextual condition $c_k$, and computed by using the songs from set $S_{c_{k}}$ that contains all $|S_{c_k}|$ positively interacted songs in condition $c_k$. $\Vec{GM_{c_{k}}}$ is simply the centroid of the vectors of the all songs $s \in S_{c_k}$. If, for example, the 3 audio features of $energy$, $tempo$ and $acousticness$ are used to model user preferences, we could have $\Vec{GM_{c_{1}}} = morning = [0.32, 0.44, 0.82]$ and $\Vec{GM_{c_{2}}} = evening = [0.78, 0.66, 0.21]$. These example models tell us that users generally prefer low energy and tempo songs and high acoustic songs in the morning compared to higher energy and tempo, but lower acousticness in the evening.

%\begin{equation}
%    \bm{GM_{c_{k}}} = \begin{bmatrix}\overline{a}_1 & \overline{a}_2 & ... & \overline{a}_n \end{bmatrix},\ \overline{a}_x = \frac{1}{m}\sum_{j=1}^ma_{x,s_j}\in S_{c_{k}}
%    \label{eq:global_model_formula}
%\end{equation}

%where $\bm{GM_{c_{k}}}$ is a vector representing the global model for contextual condition $c_k$, $\overline{a}_x$ represents the average value of audio feature $x$, $a_{x,s_j}$ represents audio feature $x$ for song $j$ from set $S_{c_{k}}$ that contains all $m$ positively interacted songs in condition $c_k$. If, for example, the 3 audio features of $energy$, $tempo$ and $acousticness$ are used to model user preferences, we could have $\bm{GM_{c_{1}}} = morning = [0.32, 0.44, 0.82]$ and $\bm{GM_{c_{2}}} = evening = [0.78, 0.66, 0.21]$. These example models tell us that users generally prefer low energy and tempo songs and high acoustic songs in the morning compared to higher energy and tempo but lower acousticness in the evening.

\subsection{Personalized model}\label{subsection:personal_model}
The personalized model is broadly comparable to the global model. Also here, audio features are used to represent user preferences for specific contextual conditions. The personalized model, however, separately creates preference models for each user instead of creating a global model based on all positive user interactions. It is represented as follows:

\begin{equation}
    \Vec{PM_{c_{k},u}} = \left[a_1, a_2, \ldots, a_n\right] = \frac{1}{|S_{c_k,u}|} \cdot \sum_{s_j \in S_{c_k,u}} \Vec{s_j},
    \label{eq:personal_model_formula}
\end{equation}

%\begin{equation}
 %   \bm{PM_{c_{k},u_{b}}} = \begin{bmatrix}\overline{a}_1 & \overline{a}_2 & ... & \overline{a}_n \end{bmatrix},\  \overline{a}_x = \frac{1}{m}\sum_{j=1}^ma_{x,s_j}\in S_{c_{k},u_{b}}
 %   \label{eq:personal_model_formula}
%\end{equation}

\noindent where $\Vec{PM_{c_{k},u}}$ is a vector representing the personalized model for contextual condition $c_k$ and user $u \in U$, and computed by using the songs from set $S_{c_k,u}$ that contains all $|S_{c_k,u}|$ positively interacted songs in condition $c_k$ by user $u$. As an example we use the 3 audio features of $energy$, $tempo$ and $acousticness$ again for user 1 and user 2. User 1 prefers to listen to calm piano music during breakfast, while user 2 likes to use energetic dance music to get more awake in the morning. Possible personalized models would be $\Vec{PM_{c_{1},u_{1}}} = morning, \; user \; 1 \; = [0.1, 0.16, 0.98]$ and $\Vec{PM_{c_{1},u_{2}}} = morning, \; user \; 2 \; = [0.88, 0.76, 0.18]$.

%where $\bm{PM_{c_{k},u_{b}}}$ is a vector representing the personalized model for contextual condition $c_k$ and user $u_b$. $\overline{a}_x$ represents the average value of audio feature $x$, $a_{x,s_j}$ represents audio feature $x$ for song $j$ from set $S_{c_{k},u_{b}}$ that contains all $m$ positively interacted songs of user $u_b$ in contextual condition $c_k$. As an example we use the 3 audio features of $energy$, $tempo$ and $acousticness$ again for Person 1 and Person 2. Person 1 prefers to listen to calm piano music during breakfast, while Person 2 likes to use energetic dance music to get more awake in the morning. Possible personalized models would be $\bm{PM_{c_{1},u_{1}}} = Person \; 1, \; morning \; = [0.1, 0.16, 0.98]$ and $\bm{PM_{c_{1},u_{2}}} = Person \; 2, \; morning \; = [0.88, 0.76, 0.18]$.

The personalized model is a more fine grained and computationally expensive approach than the global model. For each unique user and contextual condition, a specific preference model is created. These models can be seen as user profiles, representing user specific music preferences using audio features. The idea is that various conditions have different influences on the preferences of different users. Each model specifically represents this preference for a given user, in contrast to the global model, which assumes a general preference across all users. 

\subsection{Re-Ranking Score Calculation}\label{subsection:rerank_scoring}
The next step is re-ranking songs in a given recommendation list (generated by an initial recommendation algorithm) based on the similarity between their audio features vector and the audio features vector of the given contextual condition. The idea is to re-rank songs that are similar to what users like in a specific condition to higher positions in the recommendation list. In the same way, songs that are less similar will be given a lower position. The resulting scoring function applied to each song is as follows:

\begin{equation}
    new\_score = \lambda * Sim(\Vec{s_j},\Vec{GM_{c_{k}}}) + (1-\lambda) * Rec(u,s_j,c_k)',
    \label{eq:final_re_rank_calculation}
\end{equation}

\noindent where $\lambda$ is a balancing parameter, ranging from 0 to 1, that let us control the weight of the contextual similarity score relative to the weight of the initial recommendation score, $Sim(\Vec{s_j},\Vec{GM_{c_{k}}})$ is the similarity between the audio features vector of song $j$, $\Vec{s_j}$, and the audio feature vector, which can be either the global or personalized model representation, ($\Vec{GM_{c_{k}}}$ or $\Vec{PM_{c_{k},u}}$), of contextual condition $c_k$ and $Rec(u,s_j,c_k)'$ is the unity-based normalized recommendation score for user $u$ generated by an initial recommender, song $s_j$ and condition $c_k$ if a context-aware recommender is used. 

Since the audio feature representations are basically multidimensional vectors, we can use the unity-based normalized Euclidean distance ($d'$) as part of the similarity measurement. It is defined as follows:

\begin{equation}
    d(\Vec{s_j},\Vec{GM_{c_{k}}})' = \sqrt{\sum_{i=1}^n(a_{i,s_j}-GM_{c_{k},a_{i}})^2},
    \label{eq:euclidean_distance}
\end{equation}

\noindent where $a_{i,s_j}$ represents audio feature $a_i$ for song $s_j$ and $GM_{c,a_{i}}$ is the average audio feature $a_i$ within the global (or personalized) model for contextual condition $c_k$, as calculated by either Equation~\ref{eq:global_model_formula} or~\ref{eq:personal_model_formula}. This gives us the distance between the song and condition, so in order to obtain a similarity value between 0 and 1 we use:

\begin{equation}
    Sim(\Vec{s_j},\Vec{GM_{c_{k}}}) = 1 - d(\Vec{s_j},\Vec{GM_{c_{k}}})'.
    \label{eq:rerank_similarity}
\end{equation}

This gives us the freedom to replace $d'$ with any other distance metric. In addition to the regular re-ranking score calculation, we define an opposite re-ranking score calculation. It uses the same equation as Equation~\ref{eq:final_re_rank_calculation}, except that the similarity is replaced by the unity-based normalized Euclidean distance. This results in:

\begin{equation}
    opposite\_score = \lambda * d(\Vec{s_j},\Vec{GM_{c_{k}}})' + (1-\lambda) * Rec(u,s_j,c_k)',
    \label{eq:final_opposite_re_rank_calculation}
\end{equation}

\noindent where all variables are the same as described above. This opposite algorithm gives higher ranks to songs that are more different from the given context measured over the audio feature vectors. We include this to evaluate the value of re-ranking independent of the audio features vector similarity measurement. A decrease in recommendation accuracy shows a correlation between our proposed re-ranking and final recommendation quality.

\section{Experiment}\label{section:experiment}
\begin{figure}[ht]
\captionsetup{justification=centering}
\centering
\scalebox{0.5}{
      \includegraphics{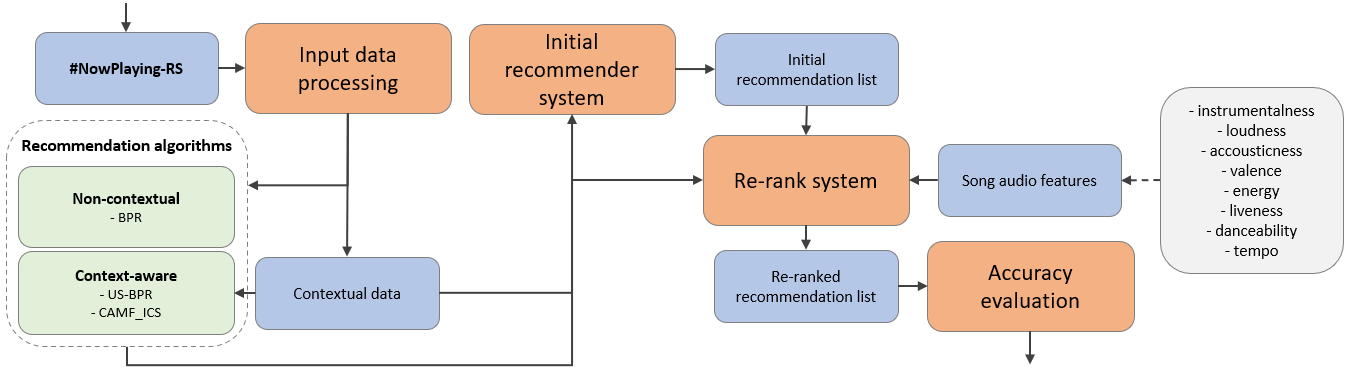}
    }
  \setlength{\belowcaptionskip}{-10pt}
  \caption{An overview of the full experimental pipeline.}
  \label{fig:experiment_pipeline}
\end{figure}

We describe the implementation and evaluation of the re-ranking algorithm in this section. First, an elaboration on the used dataset is given followed by a discussion on the initial recommendation algorithms. Thereafter, we discuss our re-ranking implementation and the obtained results. The full pipeline of the experiment is shown in Figure~\ref{fig:experiment_pipeline}.

\subsection{Datasets}\label{subsection:input_data}
For our experiment we use the \#NowPlaying-RS~\cite{poddar2018nowplaying} and InCarMusic~\cite{baltrunas2011incarmusic} datasets. However, initial results for the InCarMusic dataset are strongly inconsistent due to its sparsity. For this reason we decide to continue only with the \#NowPlaying-RS dataset. The partial InCarMusic results can be found in the re-ranking system GitHub repository. 

\#NowPlaying-RS is a comprehensive implicit feedback dataset consisting of user-song interactions crawled from Twitter and enhanced with audio features from Spotify. We assume that users listening to and tweeting about songs represent positive interactions. This dataset is rich in interactions, but limited in contextual dimensions. Only the \textit{time of day} dimension is consistently represented in the dataset. So in our experiment we only evaluate this dimension, despite the model being context agnostic. To create a manageable subset, we remove all user who have less than 3000 interactions and songs that are listened to less than 200 times. This results in a subset of 7304 songs, 333 users and 108,202 listening events. Furthermore, we categorize each interaction to the \textit{morning}, \textit{afternoon}, \textit{evening}, or \textit{night} condition based on the user's local interaction time. 

\subsection{Initial Recommendation Algorithms}\label{subsection:initial_recommendation}
To apply the re-ranking algorithm, an initial recommendation list is needed as input. We use the CARSKit by Zheng et al.~\cite{zheng2015carskit} to create such recommendation lists by using training sets. We modified the system to our own needs.\footnote{The modified version of CARSKit is accessible at \url{https://github.com/boninggong/CARSKitModified}} Next to this, we use a simple 5-fold cross validation, where 80\% of the interactions are used as training set and the 20\% of the interactions used as test set, to reduce variance and bias in the results. The cross validation outputs various training and test sets, where the songs of user-condition-song interactions are considered as relevant songs. We use the following recommendation algorithms to generate initial recommendation lists:

\begin{itemize}
  \item \textbf{Bayesian Probability Ranking (BPR)}: A simple ranking algorithm based on Bayesian probabilities~\cite{rendle2012bpr}.
  \item \textbf{UserSplitting-BPR (US-BPR)}: A contextual pre-filter algorithm that splits users in sub profiles based on contextual conditions before running the BPR algorithm~\cite{said2011inferring}.
  \item \textbf{Context-Aware Matrix Factorization - Independent Context Similarity (CAMF\_ICS)}: A specific type of ranking based matrix factorization that takes context into account based on underlying similarities of conditions within the same dimension~\cite{zheng2015similarity}.
\end{itemize}

We use ranking based recommendation algorithms here, because of the implicit nature of the \#NowPlaying-RS dataset. Next to this, we have a mix of traditional and context-aware recommender algorithms. The reason to also re-rank recommendations that are created using contextual information is to evaluate the impact of audio features based contextual re-ranking. For each output recommendation list (initial recommendation lists generated by one of the BPR, US-BPR and CAMF\_ICS), we take the top 200, 100, 50 and 25 songs as input for the re-ranking algorithm. Due to the page limit we only show results for the top 50 songs in this paper. 

Our implementation and an overview of all results can be found in our external appendix. \footnote{\url{https://github.com/boninggong/Re-rankSystem}}

\subsection{Re-Ranking Procedure}\label{subsection:rerank_procedure}
\begin{figure}[ht]
\captionsetup{justification=centering}
\centering
\scalebox{0.41}{
      \includegraphics{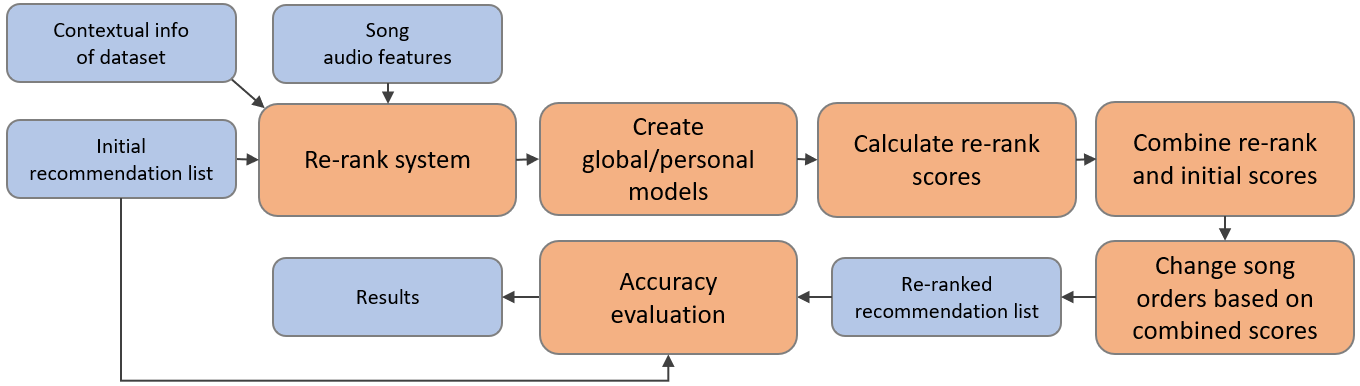}
    }
  \setlength{\belowcaptionskip}{-10pt}
  \caption{An overview of the steps carried out in the re-ranking system.}
  \label{fig:rerank_system}
\end{figure}

All steps in our re-ranking implementation are depicted in Figure~\ref{fig:rerank_system}. The recommendation lists, training data sets and audio features of songs are used as input. The system proceeds to build both global and personalized models, according to Eqs.~\ref{eq:global_model_formula} and~\ref{eq:personal_model_formula} respectively. Based on these models, it goes through each song for each recommendation list to calculate the similarity to the given contextual condition. This score is combined with the original recommendation score through the balancing factor $\lambda$. We test all $\lambda$ values between 0 and 1 with steps of 0.1. Songs are then re-ranked in a descending order based on their new scores. Lastly, the re-ranked recommendation list for both global and personalized models are evaluated against the initial recommendation lists by using the test sets.

\subsection{Evaluation Metrics}\label{subsection:evaluation_metrics}
We compare both global and personalized model based re-ranked recommendation lists to each other and the initial recommendation list. Since we try to give relevant songs a higher rank, we use ranking based accuracy metrics to evaluate the results. For the evaluation, we use precision at position $k$ (Prec@$k$) and mean average precision at position $k$ (MAP@$k$). We only show the results of the MAP@$k$ metric in this paper, because of the page limit. The results for the straightforward and widely used Prec@$k$ metric can be found in our \textit{Re-rankSystem} GitHub repository. Before defining MAP@$k$, we need to define average precision at $k$ (AP@$k$) first, which is defined as follows:

\begin{equation}
    \text{AP}@k(L) = \frac{\sum_{k=1}^m\frac{\#\{rel\;\;songs\;\;in\;\;top\;\;k\}}{k}*b_k}{\#\{rel\;\;songs\}},
    \label{eq:average_precision_at_k}
\end{equation}

\noindent where $m$ represents the total amount of items for a given recommendation list $L$ and $b_k$ is a binary value for whether the item is relevant (1) or not (0). MAP@$k$ can then be defined as the mean value of all average precision values measured over the top $k$ items over all recommendation lists. 

\begin{figure}[ht]
\captionsetup{justification=centering}
\centering
\scalebox{0.75}{
        \includegraphics[width=1\linewidth]{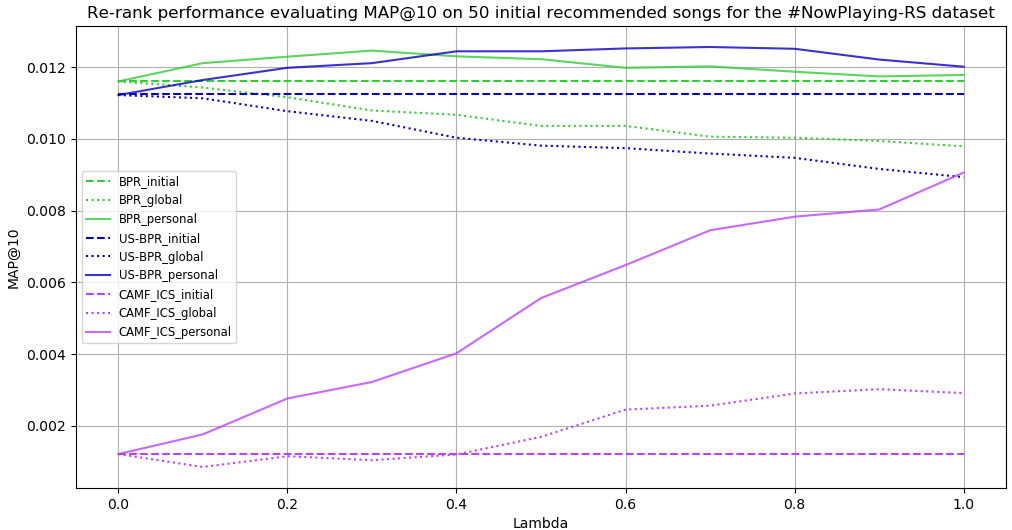}
    }
    \setlength{\belowcaptionskip}{-5pt}
    \caption{Line plot of the MAP@10 evaluation for re-ranking the top 50 songs on the \#NowPlaying-RS dataset.}
    \label{fig:map_10_plot}
\end{figure}

\subsection{Results}\label{subsection:rerank_results}
 Figure~\ref{fig:map_10_plot} visualizes the selected results for the MAP@10 metric for all three initial recommendation algorithms (BPR, US-BPR and CAMF\_IC),  comparing the global and personalized models. The first observation is that BPR and US-BPR both significantly outperform CAMF\_ICS. Another observation is that the recommendation accuracy of the re-ranked personalized model generally outperforms the accuracy of the initial (original rank) recommendations, especially for the CAMF\_ICS algorithm. The personalized model, furthermore, consistently outperforms the global model, which varies widely for each initial recommender algorithm.

\begin{figure}[ht]
\captionsetup{justification=centering}
\centering
\scalebox{0.75}{
        \includegraphics[width=1\linewidth]{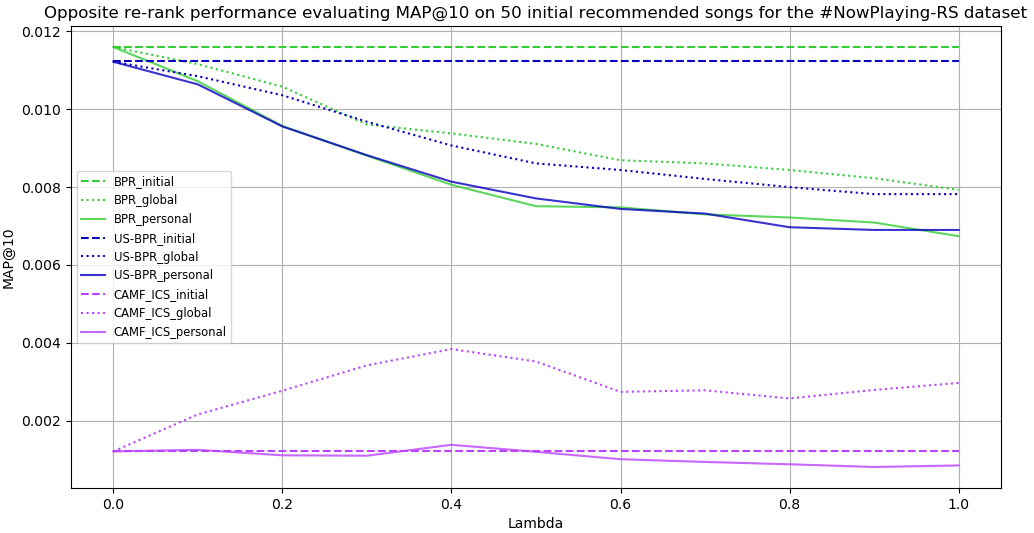}
    }
    \setlength{\belowcaptionskip}{-5pt}
    \caption{Line plot of the MAP@10 evaluation for opposite re-ranking the top 50 songs on the \#NowPlaying-RS dataset.}
    \label{fig:map_10_plot_op}
\end{figure}

To investigate the benefit of re-ranking, independent of the audio features vector similarity measurement, we measure MAP@10 for the re-ranking results using inverse ranking, as shown in Figure~\ref{fig:map_10_plot_op}. We observe that the personalized model greatly decreases accuracy compared to the initial recommendation. This strengthens the value of the re-ranking approach used.
%since re-ranking using difference measurements yields decreased accuracy.
The global model decreases accuracy for the BPR and US-BPR algorithms compared to the initial recommendation, but, surprisingly, increases for CAMF\_ICS. One possible reason is that CAMF\_ICS (in comparison to the two others) recommends fewer relevant songs overall, as well as fewer relevant songs in the top 10.

There is no single $\lambda$ value that consistently gives the best re-ranking results, even for the well performing personalized model based re-ranking results. The optimal value differs per underlying recommendation algorithm. This means that if this re-ranking algorithm would be implemented in practice, $\lambda$ should be optimized in relation to the underlying recommender algorithm. The complete results are available in the \textit{Re-rankSystem} repository.

% Prec@$k$ is defined as follows:

% \begin{equation}
%     \text{Prec}@k(L) = \frac{\#\{relevant\;\;songs\;\;in\;\;top\;\;k\;\;recommendations\}}{k}
%     \label{eq:precision_at_k}
% \end{equation}

% where $L$ represents a recommendation list and $k$ can be any number smaller than the length of $L$. So Prec@$k$ measures relatively how many relevant test set songs there are in the top $k$ songs. After having this definition, the average precision at $k$ is defined as:

% \begin{figure}[h]
% \centering
% \begin{subfigure}{.5\columnwidth}
%     \centering
%     \includegraphics[width=0.99\linewidth]{images/Prec_10.png}
%     \caption{Using the regular re-rank score.}
%     %\label{subfig:activity_radar_plot}
% \end{subfigure}%
% \begin{subfigure}{.5\columnwidth}
%     \centering
%     \includegraphics[width=0.99\linewidth]{images/Prec_10_op.png}
%     \caption{Using the opposite re-rank score.}
%     %\label{subfig:mood_radar_plot}
% \end{subfigure}
% \setlength{\belowcaptionskip}{-15pt}
% \caption{Line plots visualizing the Prec@10 evaluation for the top 50 songs on the \#NowPlaying-RS dataset.}
% \label{fig:rerank_res_prec_10}
% \end{figure}
\section{Conclusions and Future Work}\label{section:conclusion}
In this paper, first, we showed that there exists a significant correlation between audio features and  contextual conditions. Thus, audio features can be used to distinguish between different conditions when listening to music. From this, we proposed a contextual re-ranking algorithm that utilizes audio feature representations of user preferences for specific contextual conditions to re-rank any given recommendation list. Two user preference representations were presented, a global and a personalized contextual model. We evaluated the re-ranking using the two models on several recommendation algorithms' output. This was done using the \#NowPlaying-RS dataset and accuracy has been measured using the Prec@$k$ and MAP@$k$ metrics. Initial results on the contextual dimension \textit{time of day} show there is merit in applying our re-ranking algorithm. Especially the personalized model shows promising results and consistently outperforms the global model (which in turn improves the non-contextual initial recommendations). 

In future work we plan to further improve the two user preference models and re-ranking scoring. This can be done by weighing the audio features or only using a selection of them or through different similarity measurements. Furthermore, we plan to develop the audio feature representation of user preferences to include multiple contextual conditions and evaluate the resulting impact. Moreover, next to the offline evaluation, carrying out an online evaluation with actual users will provide valuable insights. 

% \begin{acks}

% \end{acks}

\bibliographystyle{ACM-Reference-Format}
\bibliography{references}


\begin{thebibliography}{29}

%%% the Web address.

\ifx \showCODEN    \undefined \def \showCODEN     #1{\unskip}     \fi
\ifx \showDOI      \undefined \def \showDOI       #1{#1}\fi
\ifx \showISBNx    \undefined \def \showISBNx     #1{\unskip}     \fi
\ifx \showISBNxiii \undefined \def \showISBNxiii  #1{\unskip}     \fi
\ifx \showISSN     \undefined \def \showISSN      #1{\unskip}     \fi
\ifx \showLCCN     \undefined \def \showLCCN      #1{\unskip}     \fi
\ifx \shownote     \undefined \def \shownote      #1{#1}          \fi
\ifx \showarticletitle \undefined \def \showarticletitle #1{#1}   \fi
\ifx \showURL      \undefined \def \showURL       {\relax}        \fi
% invisible to TeX
\providecommand\bibfield[2]{#2}
\providecommand\bibinfo[2]{#2}
\providecommand\natexlab[1]{#1}
\providecommand\showeprint[2][]{arXiv:#2}

\bibitem[\protect\citeauthoryear{Adomavicius, Sankaranarayanan, Sen, and
  Tuzhilin}{Adomavicius et~al\mbox{.}}{2005}]%
        {adomavicius2005incorporating}
\bibfield{author}{\bibinfo{person}{Gediminas Adomavicius},
  \bibinfo{person}{Ramesh Sankaranarayanan}, \bibinfo{person}{Shahana Sen},
  {and} \bibinfo{person}{Alexander Tuzhilin}.} \bibinfo{year}{2005}\natexlab{}.
\newblock \showarticletitle{Incorporating contextual information in recommender
  systems using a multidimensional approach}.
\newblock \bibinfo{journal}{\emph{ACM Transactions on Information Systems
  (TOIS)}} \bibinfo{volume}{23}, \bibinfo{number}{1} (\bibinfo{year}{2005}),
  \bibinfo{pages}{103--145}.
\newblock


\bibitem[\protect\citeauthoryear{Adomavicius and Tuzhilin}{Adomavicius and
  Tuzhilin}{2011}]%
        {adomavicius2011context}
\bibfield{author}{\bibinfo{person}{Gediminas Adomavicius} {and}
  \bibinfo{person}{Alexander Tuzhilin}.} \bibinfo{year}{2011}\natexlab{}.
\newblock \showarticletitle{Context-aware recommender systems}.
\newblock In \bibinfo{booktitle}{\emph{Recommender systems handbook}}.
  \bibinfo{publisher}{Springer}, \bibinfo{pages}{217--253}.
\newblock


\bibitem[\protect\citeauthoryear{Andjelkovic, Parra, and O'Donovan}{Andjelkovic
  et~al\mbox{.}}{2016}]%
        {andjelkovic2016moodplay}
\bibfield{author}{\bibinfo{person}{Ivana Andjelkovic}, \bibinfo{person}{Denis
  Parra}, {and} \bibinfo{person}{John O'Donovan}.}
  \bibinfo{year}{2016}\natexlab{}.
\newblock \showarticletitle{Moodplay: Interactive mood-based music discovery
  and recommendation}. In \bibinfo{booktitle}{\emph{Proceedings of the 2016
  Conference on User Modeling Adaptation and Personalization}}.
  \bibinfo{pages}{275--279}.
\newblock


\bibitem[\protect\citeauthoryear{Baltrunas and Amatriain}{Baltrunas and
  Amatriain}{2009}]%
        {baltrunas2009towards}
\bibfield{author}{\bibinfo{person}{Linas Baltrunas} {and}
  \bibinfo{person}{Xavier Amatriain}.} \bibinfo{year}{2009}\natexlab{}.
\newblock \showarticletitle{Towards time-dependant recommendation based on
  implicit feedback}. In \bibinfo{booktitle}{\emph{Workshop on context-aware
  recommender systems (CARS’09)}}. Citeseer, \bibinfo{pages}{25--30}.
\newblock


\bibitem[\protect\citeauthoryear{Baltrunas, Kaminskas, Ludwig, Moling, Ricci,
  Aydin, L{\"u}ke, and Schwaiger}{Baltrunas et~al\mbox{.}}{2011}]%
        {baltrunas2011incarmusic}
\bibfield{author}{\bibinfo{person}{Linas Baltrunas}, \bibinfo{person}{Marius
  Kaminskas}, \bibinfo{person}{Bernd Ludwig}, \bibinfo{person}{Omar Moling},
  \bibinfo{person}{Francesco Ricci}, \bibinfo{person}{Aykan Aydin},
  \bibinfo{person}{Karl-Heinz L{\"u}ke}, {and} \bibinfo{person}{Roland
  Schwaiger}.} \bibinfo{year}{2011}\natexlab{}.
\newblock \showarticletitle{Incarmusic: Context-aware music recommendations in
  a car}. In \bibinfo{booktitle}{\emph{Conference on electronic commerce and
  web technologies}}. Springer, \bibinfo{pages}{89--100}.
\newblock


\bibitem[\protect\citeauthoryear{Braunhofer, Kaminskas, and Ricci}{Braunhofer
  et~al\mbox{.}}{2013}]%
        {braunhofer2013location}
\bibfield{author}{\bibinfo{person}{Matthias Braunhofer},
  \bibinfo{person}{Marius Kaminskas}, {and} \bibinfo{person}{Francesco Ricci}.}
  \bibinfo{year}{2013}\natexlab{}.
\newblock \showarticletitle{Location-aware music recommendation}.
\newblock \bibinfo{journal}{\emph{Journal of Multimedia Information Retrieval}}
  \bibinfo{volume}{2}, \bibinfo{number}{1} (\bibinfo{year}{2013}),
  \bibinfo{pages}{31--44}.
\newblock


\bibitem[\protect\citeauthoryear{Chen, Tsai, Liu, and Yang}{Chen
  et~al\mbox{.}}{2013}]%
        {chen2013using}
\bibfield{author}{\bibinfo{person}{Chih-Ming Chen}, \bibinfo{person}{Ming-Feng
  Tsai}, \bibinfo{person}{Jen-Yu Liu}, {and} \bibinfo{person}{Yi-Hsuan Yang}.}
  \bibinfo{year}{2013}\natexlab{}.
\newblock \showarticletitle{Using emotional context from article for contextual
  music recommendation}. In \bibinfo{booktitle}{\emph{Conference on
  Multimedia}}. \bibinfo{pages}{649--652}.
\newblock


\bibitem[\protect\citeauthoryear{Cheng and Shen}{Cheng and Shen}{2014}]%
        {cheng2014just}
\bibfield{author}{\bibinfo{person}{Zhiyong Cheng} {and} \bibinfo{person}{Jialie
  Shen}.} \bibinfo{year}{2014}\natexlab{}.
\newblock \showarticletitle{Just-for-me: an adaptive personalization system for
  location-aware social music recommendation}. In
  \bibinfo{booktitle}{\emph{Conference on multimedia retrieval}}.
  \bibinfo{pages}{185--192}.
\newblock


\bibitem[\protect\citeauthoryear{Cremonesi, Garza, Quintarelli, and
  Turrin}{Cremonesi et~al\mbox{.}}{2011}]%
        {cremonesi2011top}
\bibfield{author}{\bibinfo{person}{Paolo Cremonesi}, \bibinfo{person}{Paolo
  Garza}, \bibinfo{person}{Elisa Quintarelli}, {and} \bibinfo{person}{Roberto
  Turrin}.} \bibinfo{year}{2011}\natexlab{}.
\newblock \showarticletitle{Top-n recommendations on unpopular items with
  contextual knowledge}. In \bibinfo{booktitle}{\emph{2011 Workshop on
  Context-aware Recommender Systems. Chicago}}.
\newblock


\bibitem[\protect\citeauthoryear{Cunningham, Bainbridge, and
  Falconer}{Cunningham et~al\mbox{.}}{2006}]%
        {cunningham2006more}
\bibfield{author}{\bibinfo{person}{Sally~Jo Cunningham}, \bibinfo{person}{David
  Bainbridge}, {and} \bibinfo{person}{Annette Falconer}.}
  \bibinfo{year}{2006}\natexlab{}.
\newblock \showarticletitle{"More of an art than a science": Supporting the
  creation of playlists and mixes}. In \bibinfo{booktitle}{\emph{Seventh
  International Conference on Music Information Retrieval}} (Victoria, Canada).
  \bibinfo{publisher}{University of Victoria}, \bibinfo{pages}{8--12}.
\newblock


\bibitem[\protect\citeauthoryear{Dias, Fonseca, and Cunha}{Dias
  et~al\mbox{.}}{2014}]%
        {dias2014user}
\bibfield{author}{\bibinfo{person}{Ricardo Dias}, \bibinfo{person}{Manuel~J
  Fonseca}, {and} \bibinfo{person}{Ricardo Cunha}.}
  \bibinfo{year}{2014}\natexlab{}.
\newblock \showarticletitle{A User-centered Music Recommendation Approach for
  Daily Activities}. In \bibinfo{booktitle}{\emph{CBRecSys@ RecSys}}.
  \bibinfo{pages}{26--33}.
\newblock


\bibitem[\protect\citeauthoryear{Haruna, Ismail, Damiasih, Chiroma, and
  Herawan}{Haruna et~al\mbox{.}}{2017}]%
        {haruna2017comprehensive}
\bibfield{author}{\bibinfo{person}{Khalid Haruna},
  \bibinfo{person}{Maizatul~Akmar Ismail}, \bibinfo{person}{Damiasih Damiasih},
  \bibinfo{person}{Haruna Chiroma}, {and} \bibinfo{person}{Tutut Herawan}.}
  \bibinfo{year}{2017}\natexlab{}.
\newblock \showarticletitle{Comprehensive survey on comparisons across
  contextual pre-filtering, contextual post-filtering and contextual modelling
  approaches}.
\newblock \bibinfo{journal}{\emph{Telkomnika}} \bibinfo{volume}{15},
  \bibinfo{number}{4} (\bibinfo{year}{2017}), \bibinfo{pages}{1865--1875}.
\newblock


\bibitem[\protect\citeauthoryear{Kaminskas and Ricci}{Kaminskas and
  Ricci}{2012}]%
        {kaminskas2012contextual}
\bibfield{author}{\bibinfo{person}{Marius Kaminskas} {and}
  \bibinfo{person}{Francesco Ricci}.} \bibinfo{year}{2012}\natexlab{}.
\newblock \showarticletitle{Contextual music information retrieval and
  recommendation: State of the art and challenges}.
\newblock \bibinfo{journal}{\emph{Computer Science Review}}
  \bibinfo{volume}{6}, \bibinfo{number}{2-3} (\bibinfo{year}{2012}),
  \bibinfo{pages}{89--119}.
\newblock


\bibitem[\protect\citeauthoryear{Lamche, R{\"o}dl, Hauptmann, and
  W{\"o}rndl}{Lamche et~al\mbox{.}}{2015}]%
        {lamche2015context}
\bibfield{author}{\bibinfo{person}{B{\'e}atrice Lamche},
  \bibinfo{person}{Yannick R{\"o}dl}, \bibinfo{person}{Claudius Hauptmann},
  {and} \bibinfo{person}{Wolfgang W{\"o}rndl}.}
  \bibinfo{year}{2015}\natexlab{}.
\newblock \showarticletitle{Context-Aware Recommendations for Mobile Shopping}.
  In \bibinfo{booktitle}{\emph{LocalRec@ RecSys}}. \bibinfo{pages}{21--27}.
\newblock


\bibitem[\protect\citeauthoryear{Melchiorre and Schedl}{Melchiorre and
  Schedl}{2020}]%
        {10.1145/3340631.3394874}
\bibfield{author}{\bibinfo{person}{Alessandro~B. Melchiorre} {and}
  \bibinfo{person}{Markus Schedl}.} \bibinfo{year}{2020}\natexlab{}.
\newblock \showarticletitle{Personality Correlates of Music Audio Preferences
  for Modelling Music Listeners}. In \bibinfo{booktitle}{\emph{Conference on
  User Modeling, Adaptation and Personalization}} (Genoa, Italy)
  \emph{(\bibinfo{series}{UMAP ’20})}. \bibinfo{publisher}{Association for
  Computing Machinery}, \bibinfo{address}{New York, NY, USA},
  \bibinfo{pages}{313–317}.
\newblock
\showISBNx{9781450368612}


\bibitem[\protect\citeauthoryear{North, Hargreaves, and Hargreaves}{North
  et~al\mbox{.}}{2004}]%
        {north2004uses}
\bibfield{author}{\bibinfo{person}{Adrian~C North}, \bibinfo{person}{David~J
  Hargreaves}, {and} \bibinfo{person}{Jon~J Hargreaves}.}
  \bibinfo{year}{2004}\natexlab{}.
\newblock \showarticletitle{Uses of music in everyday life}.
\newblock \bibinfo{journal}{\emph{Music perception}} \bibinfo{volume}{22},
  \bibinfo{number}{1} (\bibinfo{year}{2004}), \bibinfo{pages}{41--77}.
\newblock


\bibitem[\protect\citeauthoryear{Panniello, Tuzhilin, Gorgoglione, Palmisano,
  and Pedone}{Panniello et~al\mbox{.}}{2009}]%
        {panniello2009experimental}
\bibfield{author}{\bibinfo{person}{Umberto Panniello},
  \bibinfo{person}{Alexander Tuzhilin}, \bibinfo{person}{Michele Gorgoglione},
  \bibinfo{person}{Cosimo Palmisano}, {and} \bibinfo{person}{Anto Pedone}.}
  \bibinfo{year}{2009}\natexlab{}.
\newblock \showarticletitle{Experimental comparison of pre-vs. post-filtering
  approaches in context-aware recommender systems}. In
  \bibinfo{booktitle}{\emph{Proceedings of the third ACM conference on
  Recommender systems}}. \bibinfo{pages}{265--268}.
\newblock


\bibitem[\protect\citeauthoryear{Pichl, Zangerle, and Specht}{Pichl
  et~al\mbox{.}}{2016}]%
        {pichl2016understanding}
\bibfield{author}{\bibinfo{person}{Martin Pichl}, \bibinfo{person}{Eva
  Zangerle}, {and} \bibinfo{person}{G{\"u}nther Specht}.}
  \bibinfo{year}{2016}\natexlab{}.
\newblock \showarticletitle{Understanding playlist creation on music streaming
  platforms}. In \bibinfo{booktitle}{\emph{2016 IEEE International Symposium on
  Multimedia (ISM)}}. IEEE, \bibinfo{pages}{475--480}.
\newblock


\bibitem[\protect\citeauthoryear{Poddar, Zangerle, and Yang}{Poddar
  et~al\mbox{.}}{2018}]%
        {poddar2018nowplaying}
\bibfield{author}{\bibinfo{person}{Asmita Poddar}, \bibinfo{person}{Eva
  Zangerle}, {and} \bibinfo{person}{Yi-Hsuan Yang}.}
  \bibinfo{year}{2018}\natexlab{}.
\newblock \showarticletitle{\#nowplaying-RS: A New Benchmark Dataset for
  Building Context-Aware Music Recommender Systems}. In
  \bibinfo{booktitle}{\emph{Proceedings of the 15th Sound \& Music Computing
  Conference}} (2018-07-04). \bibinfo{address}{Limassol, Cyprus}.
\newblock
\newblock
\shownote{code at https://github.com/asmitapoddar/nowplaying-RS-Music-Reco-FM.}


\bibitem[\protect\citeauthoryear{Rendle, Freudenthaler, Gantner, and
  Schmidt-Thieme}{Rendle et~al\mbox{.}}{2009}]%
        {rendle2012bpr}
\bibfield{author}{\bibinfo{person}{Steffen Rendle}, \bibinfo{person}{Christoph
  Freudenthaler}, \bibinfo{person}{Zeno Gantner}, {and} \bibinfo{person}{Lars
  Schmidt-Thieme}.} \bibinfo{year}{2009}\natexlab{}.
\newblock \showarticletitle{BPR: Bayesian personalized ranking from implicit
  feedback}. In \bibinfo{booktitle}{\emph{Proceedings of the Twenty-Fifth
  Conference on Uncertainty in Artificial Intelligence}}.
  \bibinfo{pages}{452--461}.
\newblock


\bibitem[\protect\citeauthoryear{Said, De~Luca, and Albayrak}{Said
  et~al\mbox{.}}{2011}]%
        {said2011inferring}
\bibfield{author}{\bibinfo{person}{Alan Said}, \bibinfo{person}{Ernesto~W
  De~Luca}, {and} \bibinfo{person}{Sahin Albayrak}.}
  \bibinfo{year}{2011}\natexlab{}.
\newblock \showarticletitle{Inferring contextual user profiles-improving
  recommender performance}. In \bibinfo{booktitle}{\emph{Proceedings of the 3rd
  RecSys Workshop on Context-Aware Recommender Systems}}.
\newblock


\bibitem[\protect\citeauthoryear{Schedl, Flexer, and Urbano}{Schedl
  et~al\mbox{.}}{2013}]%
        {schedl2013neglected}
\bibfield{author}{\bibinfo{person}{Markus Schedl}, \bibinfo{person}{Arthur
  Flexer}, {and} \bibinfo{person}{Juli{\'a}n Urbano}.}
  \bibinfo{year}{2013}\natexlab{}.
\newblock \showarticletitle{The neglected user in music information retrieval
  research}.
\newblock \bibinfo{journal}{\emph{Journal of Intelligent Information Systems}}
  \bibinfo{volume}{41}, \bibinfo{number}{3} (\bibinfo{year}{2013}),
  \bibinfo{pages}{523--539}.
\newblock


\bibitem[\protect\citeauthoryear{Schedl and Schnitzer}{Schedl and
  Schnitzer}{2014}]%
        {schedl2014location}
\bibfield{author}{\bibinfo{person}{Markus Schedl} {and}
  \bibinfo{person}{Dominik Schnitzer}.} \bibinfo{year}{2014}\natexlab{}.
\newblock \showarticletitle{Location-aware music artist recommendation}. In
  \bibinfo{booktitle}{\emph{Conf. on multimedia modeling}}. Springer,
  \bibinfo{pages}{205--213}.
\newblock


\bibitem[\protect\citeauthoryear{Villegas, S{\'a}nchez, D{\'\i}az-Cely, and
  Tamura}{Villegas et~al\mbox{.}}{2018}]%
        {villegas2018characterizing}
\bibfield{author}{\bibinfo{person}{Norha~M Villegas}, \bibinfo{person}{Cristian
  S{\'a}nchez}, \bibinfo{person}{Javier D{\'\i}az-Cely}, {and}
  \bibinfo{person}{Gabriel Tamura}.} \bibinfo{year}{2018}\natexlab{}.
\newblock \showarticletitle{Characterizing context-aware recommender systems: A
  systematic literature review}.
\newblock \bibinfo{journal}{\emph{Knowledge-Based Systems}}
  \bibinfo{volume}{140} (\bibinfo{year}{2018}), \bibinfo{pages}{173--200}.
\newblock


\bibitem[\protect\citeauthoryear{Wang and Wang}{Wang and Wang}{2014}]%
        {wang2014improving}
\bibfield{author}{\bibinfo{person}{Xinxi Wang} {and} \bibinfo{person}{Ye
  Wang}.} \bibinfo{year}{2014}\natexlab{}.
\newblock \showarticletitle{Improving content-based and hybrid music
  recommendation using deep learning}. In \bibinfo{booktitle}{\emph{Conf. on
  Multimedia}}. \bibinfo{pages}{627--636}.
\newblock


\bibitem[\protect\citeauthoryear{Zhang, Yao, Sun, and Tay}{Zhang
  et~al\mbox{.}}{2019}]%
        {zhang2019deep}
\bibfield{author}{\bibinfo{person}{Shuai Zhang}, \bibinfo{person}{Lina Yao},
  \bibinfo{person}{Aixin Sun}, {and} \bibinfo{person}{Yi Tay}.}
  \bibinfo{year}{2019}\natexlab{}.
\newblock \showarticletitle{Deep learning based recommender system: A survey
  and new perspectives}.
\newblock \bibinfo{journal}{\emph{ACM Computing Surveys (CSUR)}}
  \bibinfo{volume}{52}, \bibinfo{number}{1} (\bibinfo{year}{2019}),
  \bibinfo{pages}{1--38}.
\newblock


\bibitem[\protect\citeauthoryear{Zheng}{Zheng}{2018}]%
        {zheng2018context}
\bibfield{author}{\bibinfo{person}{Yong Zheng}.}
  \bibinfo{year}{2018}\natexlab{}.
\newblock \showarticletitle{Context-aware mobile recommendation by a novel
  post-filtering approach}. In \bibinfo{booktitle}{\emph{The Thirty-First
  International Flairs Conference}}.
\newblock


\bibitem[\protect\citeauthoryear{Zheng, Mobasher, and Burke}{Zheng
  et~al\mbox{.}}{2015a}]%
        {zheng2015carskit}
\bibfield{author}{\bibinfo{person}{Yong Zheng}, \bibinfo{person}{Bamshad
  Mobasher}, {and} \bibinfo{person}{Robin Burke}.}
  \bibinfo{year}{2015}\natexlab{a}.
\newblock \showarticletitle{CARSKit: A Java-Based Context-Aware Recommendation
  Engine}. In \bibinfo{booktitle}{\emph{2015 IEEE International Conference on
  Data Mining Workshop (ICDMW)}}. \bibinfo{pages}{1668--1671}.
\newblock


\bibitem[\protect\citeauthoryear{Zheng, Mobasher, and Burke}{Zheng
  et~al\mbox{.}}{2015b}]%
        {zheng2015similarity}
\bibfield{author}{\bibinfo{person}{Yong Zheng}, \bibinfo{person}{Bamshad
  Mobasher}, {and} \bibinfo{person}{Robin Burke}.}
  \bibinfo{year}{2015}\natexlab{b}.
\newblock \showarticletitle{Similarity-based context-aware recommendation}. In
  \bibinfo{booktitle}{\emph{International Conference on Web Information Systems
  Engineering}}. Springer, \bibinfo{pages}{431--447}.
\newblock


\end{thebibliography}

\appendix

\end{document}